\documentclass[a4paper,10pt]{article}
\usepackage[russian,english]{babel}
\usepackage[T2A]{fontenc}
\usepackage[utf8]{inputenc}
\usepackage{amsmath}
\usepackage{amsthm}
\usepackage{amssymb}
\usepackage{textcomp}
\usepackage{array}
\usepackage{eepic}
\usepackage{graphicx}
\usepackage[usenames]{color}
\usepackage[matrix,arrow,curve]{xy}

\usepackage[unicode,pdfborder={0 0 1}]{hyperref}

\newcommand{\eqlb}[2]{\begin{equation} \label{#1} #2 \end{equation}}
\newcommand{\eq}[1]{\begin{equation*} #1 \end{equation*}}
\newcommand{\eqa}[1]{\begin{eqnarray*} #1 \end{eqnarray*}}
\newcommand{\eqal}[2]{\begin{eqnarray} \label{#1} #2 \end{eqnarray}}
\newcommand{\eqs}[1]{$ #1 $}
\newcommand{\brc}[1]{\left(#1\right)}
\newcommand{\bsq}[1]{\left[#1\right]}
\newcommand{\bfi}[1]{\left\{ #1\right\}}

\newcommand{\qq}{\qquad}
\newcommand{\wt}[1]{\widetilde{#1}}
\newcommand{\matr}[2]{\begin{array}{#1}#2\end{array}}

\newcommand{\mbe}{\mathbf{e}}
\newcommand{\blg}[1]{\left\langle #1 \right\rangle}
\newcommand{\at}[2]{\matr{c}{#1\\#2}}

\newcommand{\svs}{\\[12pt]}

\newcommand{\rmd}{\textrm{d}}
\newcommand{\rmi}{\textrm{i}}
\newcommand{\rme}{\textrm{e}}

\numberwithin{equation}{section}

\title{New $2\times2$-matrix linear problems for the Painlev\'{e} equations}
\author{G.~Aminov\\
{\sf Institute for Theoretical and Experimental Physics, Moscow, Russia}\\
{\sf Moscow Institute of Physics and Technology, Moscow, Russia}\\
{\em e-mail: aminov@itep.ru}\\
\\
S.~Arthamonov\\
{\sf Institute for Theoretical and Experimental Physics, Moscow, Russia}\\
{\sf Moscow Institute of Physics and Technology, Moscow, Russia}\\
{\em e-mail: artamonov@itep.ru}
}

\hoffset= - 1.0in
\begin{document}
\maketitle
\vspace{-9cm}
\begin{flushright}
 ITEP-TH 55/11\\
\end{flushright}
\vspace{8cm}

\begin{abstract}
We construct $2\times2$-matrix linear problems with a spectral parameter for the Painlev\'{e} equations I--V by means of the degeneration processes from the elliptic linear problem for the Painlev\'e VI equation. These processes supplement the known degeneration relations between the Painlev\'{e} equations with the degeneration
scheme for the associated linear problems. The degeneration relations constructed in this paper are based on the trigonometric, rational, and Inozemtsev limits.
\end{abstract}

\tableofcontents
\section{Introduction}

We study the Painlev\'{e} equations and the associated $2\times2$-matrix linear problems. The Painlev\'{e} equations are six nonlinear ordinary second-order differential equations discovered by P.~Painlev\'{e}, R.~Fuchs, and B.~Gambier \cite{Painleve'1900}, \cite{Painleve'1902}, \cite{Fuchs'1905}, \cite{Gambier'1906} at the beginning of the XX century. The approach to the Painlev\'{e} equations from the point of view of the monodromy preserving deformations of linear ordinary differential equations was established by Fuchs in the work \cite{Fuchs'1905} and generalized in the works by L.~Schlesinger \cite{Schlesinger'1912} and R.~Garnier \cite{Garnier'1912}, \cite{Garnier'1917}. After a long break this approach was further developed in the works \cite{FlaschkaNewel'80}, \cite{JimboMiwaUeno'1981_1}, \cite{JimboMiwa'1981_2}, \cite{JimboMiwa'1981_3}, \cite{Krichever'2002}, see also books \cite{ItsNovokshenov'86}, \cite{PainleveTrancedents'06}, \cite{Conte'99}.

Another important approach to the Painlev\'{e} equations was established in the work \cite{Malmquist'1922} --- the Hamiltonian approach. It turns out that each Painlev\'{e} equation is equivalent to the equations of motion of some non-autonomous Hamiltonian system. Such Hamiltonian systems were first introduced by K.~Okamoto in the works \cite{Okamoto'81}, \cite{Okamoto'86}, and \cite{Okamoto'80}. The next step in the development of the Hamiltonian approach was the representation of the Painlev\'{e} equations as non-autonomous Hamiltonian systems describing the motion of а particle in a nonstationary potential. The Hamiltonian of this type was constructed by Yu.~Manin in \cite{Manin'98}. Soon after A.~Levin and M.~Olshanetsky discovered \cite{LO'1997} that the Lax pair of the elliptic Calogero system forms the linear problem for the equation of isomonodromic deformations on torus and, particularly, for the Painlev\'{e} VI equation with specific choice of arbitrary constants. This connection between the Painlev\'{e} VI equation and the integrable model of the Calogero type \cite{Calogero} was called the Painlev\'{e}-Calogero correspondence. Later K.~Takasaki \cite{Takasaki'01} derived Hamiltonians for the Painlev\'{e} equations I--V from the Manin's Hamiltonian using degeneration relations (\ref{eq:pden}) between the Painlev\'{e} equations \cite{Okamoto'86}. It is worth noting that the similar diagram of degeneration (\ref{eq:pden}) was known (without any connection to the Painlev\'{e} equations) for the autonomous Inozemtsev systems \cite{Diejen'95}.
\eqlb{eq:pden}
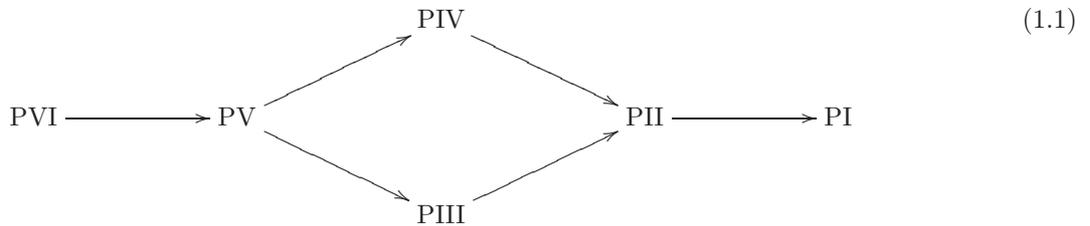
{\xymatrix{
&&&&\mathrm{PIV}\ar[drr]&&&&\\
\mathrm{PVI}\ar[rr]&&\mathrm{PV}\ar[urr]\ar[drr]&&&&\mathrm{PII}\ar[rr]&&\mathrm{PI}\\
&&&&\mathrm{PIII}\ar[urr]&&&&
}}
Thus, Takasaki extended the Painlev\'{e}-Calogero correspondence to the whole set of the Painlev\'{e} equations. Recently, this result was further developed in \cite{ZotovZabrodin'2011}, where a ``quantized'' version of the Painlev\'e-Calogero correspondence was suggested.

The goal of this paper is to construct $2\times2$-matrix linear problems with a spectral parameter for the Painlev\'{e} equations I--V by means of degeneration processes. In the work \cite{AA1} we proposed a relation between the elliptic \eqs{SL(N,\mathbb C)} top and Toda systems. This relation is based on the Inozemtsev limit \cite{Inozemtsev} and allows one to obtain the Lax pair of a Toda system from the Lax pair of the elliptic \eqs{SL(N,\mathbb C)} top. It is known that there is a connection between the systems discussed above and the Painlev\'{e} equations. The equations of motion of a non-autonomous elliptic \eqs{SL(2,\mathbb C)} top are equivalent to a particular case of the Painlev\'{e} VI equation and the equations of motion of non-autonomous Toda systems are equivalent to the Painlev\'{e} III equation with a definite choice of arbitrary constants. Thus, we can apply the procedure from \cite{AA1} to the linear problem for the Painlev\'{e} VI equation in the elliptic form \cite{Painleve'1906}, \cite{Manin'98}, \cite{BabichBordag'97}. A.~Zotov constructed the $2\times2$ Lax pair with spectral parameter \eqs{z} for the Calogero-Inozemtsev system with one degree of freedom \cite{Zotov'04}. This Lax pair also provides the linear problem for the Painlev\'{e} VI equation in the elliptic form. The Calogero-Inozemtsev system considered in \cite{Zotov'04} is described by the Hamiltonian on an elliptic curve \eqs{\blg{1,\tau}}, where the parameter \eqs{\tau} stands for the time in the non-autonomous version of this system. So, using the linear problem for the Painlev\'{e} VI equation from \cite{Zotov'04} we obtain linear problems for other Painlev\'{e} equations by means of the degeneration processes.

In Sections \ref{sec:PV} and \ref{sec:PIII} linear problems for the Painlev\'{e} equations V and III are constructed. They are obtained as limits of the linear problem for the Painlev\'{e} VI equation. The common component of these limits is the following decomposition of the parameter \eqs{\tau} of an elliptic curve:
\eq{\tau=\tau_1+\tau_2,}
where \eqs{\tau_1} stands for the time in the limiting system and \eqs{\tau_2} gives the trigonometric limit \eqs{Im\,\tau_2\rightarrow+\infty}. The difference between the limits is due to the infinite shifts of the Calogero-Inozemtsev system coordinate \eqs{u} and spectral parameter \eqs{z}. The limits also differ in the scalings of constants of the linear problem for the Painlev\'{e} VI equation \cite{Zotov'04}.

In Section \ref{V->IV} we construct a linear problem for the Painlev\'{e} IV equation using the result of Section \ref{sec:PV}. In Sections \ref{sec:PII} and \ref{sec:PI}, using the linear problem from Section \ref{sec:PIII}, we obtain linear problems for the Painlev\'{e} equations II and I, respectively. Thus, the degeneration relations between the linear problems obtained in this paper can be described by the following diagram:
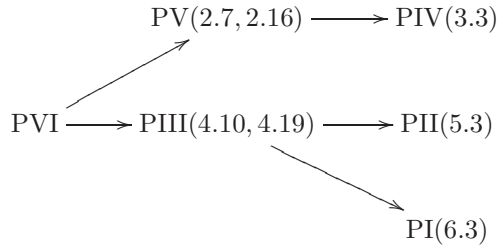
\eq{\xymatrix{
&\mathrm{PV} (\ref{eq:lax5_1}, \ref{eq:lax5_2})\ar[r]&\mathrm{PIV} (\ref{eq:lax4})\\
\mathrm{PVI}\ar[r]\ar[ru]&\mathrm{PIII} (\ref{eq:lax3_1}, \ref{eq:lax3_3})\ar[r]\ar[rd]&\mathrm{PII} (\ref{eq:lax2})\\
&&\mathrm{PI} (\ref{eq:lax1})
}}

\subsection{Painlev\'{e} equations}

We will now review general facts and notation about the equations under
consideration. The six Painlev\'{e} equations \cite{Painleve'1900}, \cite{Painleve'1902}, \cite{Fuchs'1905},
\cite{Gambier'1906} in the rational form \cite{Iwasaki'91} are
\eqa{&\textrm{PVI:}\qq\dfrac{\rmd^2\lambda}{\rmd t^2}=\dfrac12\brc{\dfrac1{\lambda}+\dfrac1{\lambda-1}+\dfrac1{\lambda-t}}\brc{\dfrac{\rmd\lambda}{\rmd t}}^2-\brc{\dfrac1t+\dfrac1{t-1}+\dfrac1{\lambda-t}}\dfrac{\rmd\lambda}{\rmd t}+\\
&+\dfrac{\lambda\brc{\lambda-1}\brc{\lambda-t}}{t^2\brc{t-1}^2}\bsq{\alpha-\beta\dfrac t{\lambda^2}+\gamma\dfrac{t-1}{\brc{\lambda-1}^2}+\brc{\dfrac12-\delta}\dfrac{t\brc{t-1}}{\brc{\lambda-t}^2}},}
\eq{\textrm{PV:}\qq\dfrac{\rmd^2\lambda}{\rmd t^2}=\brc{\dfrac1{2\lambda}+\dfrac1{\lambda-1}}\brc{\dfrac{\rmd\lambda}{\rmd t}}^2-\dfrac1t\dfrac{\rmd\lambda}{\rmd t}+\dfrac{\brc{\lambda-1}^2}{t^2}\brc{\alpha\lambda+\dfrac{\beta}{\lambda}}+\gamma\dfrac{\lambda}t+\delta\dfrac{\lambda\brc{\lambda+1}}{\brc{\lambda-1}},}
\eq{\textrm{PIV:}\qq\dfrac{\rmd^2\lambda}{\rmd t^2}=\dfrac1{2\lambda}\brc{\dfrac{\rmd\lambda}{\rmd t}}^2+\dfrac32\lambda^3+4t\lambda^2+2\brc{t^2-\alpha}\lambda+\dfrac{\beta}{\lambda},}
\eq{\textrm{PIII:}\qq\dfrac{\rmd^2\lambda}{\rmd t^2}=\dfrac1{\lambda}\brc{\dfrac{\rmd\lambda}{\rmd t}}^2-\dfrac1t\dfrac{\rmd\lambda}{\rmd t}+\dfrac1t\brc{\alpha\lambda^2+\beta}+\gamma\lambda^3+\dfrac{\delta}{\lambda},}
\eq{\textrm{PII:}\qq\dfrac{\rmd^2\lambda}{\rmd t^2}=2\lambda^3+t\lambda+\alpha,}
\eq{\textrm{PI:}\qq\dfrac{\rmd^2\lambda}{\rmd t^2}=6\lambda^2+t,}
where \eqs{\alpha, \beta, \gamma, \delta} are arbitrary complex constants. The Painlev\'{e} equations
I--V can be derived from the Painlev\'{e} VI equation by means of the degeneration processes (\ref{eq:pden})
\cite{Okamoto'86}.

\subsection{Elliptic linear problem for the Painlev\'{e} VI equation}
The sixth Painlev\'{e} equation in the elliptic form \cite{Painleve'1906}, \cite{Manin'98}, \cite{BabichBordag'97} is
\eq{\dfrac{\rmd^2 u}{\rmd \tau^2}=-2\sum_{\alpha=0}^3\nu_{\alpha}^2 E_2'\brc{u+\omega_{\alpha},\tau},}
where \eqs{\omega_{\alpha}=\bfi{0,\frac12,\frac{\tau}2,\frac{1+\tau}2}} and the second Eisenstein function \eqs{E_2(z)} (\ref{eq:eisdef}) is defined on a complex torus \eqs{\blg{1,\tau}} (see Appendix \ref{app:ellipticfunctions}). In this form the Painlev\'{e} VI equation is equivalent to the equation of motion of a non-autonomous Calogero-Inozemtsev system with one degree of freedom which is described by the Hamiltonian
\eqlb{eq:ham6}{H^{\mathrm{VI}}=v^2-\sum_{\alpha=0}^3\nu_{\alpha}^2E_2\brc{u+\omega_{\alpha}}.}
We will also need the following expression to calculate the limits in Sections \ref{V->IV}, \ref{sec:PII}, \ref{sec:PI}:
\eqlb{eq:ham6z}{H^{\mathrm{VI}}\cong H^{\mathrm{VI}}-F(z)=
v^2-\sum_{\alpha=0}^3\nu_{\alpha}^2\brc{E_2\brc{u+\omega_{\alpha}}-E_2\brc{z+\omega_{\alpha}}}.}

The non-autonomous version of the Calogero-Inozemtsev  system with one degree of freedom has the following
\eqs{2\times2} Lax representation constructed in \cite{Zotov'04}:
\eqlb{eq:idp6}{\partial_{\tau}L^{\mathrm{VI}}-\dfrac1{2\pi\rmi}\partial_{z}M^{\mathrm{VI}}=
\bsq{L^{\mathrm{VI}},M^{\mathrm{VI}}}.}
The Lax pair \eqs{L^{\mathrm{VI}}}, \eqs{M^{\mathrm{VI}}} is of the form
\begin{subequations}
\label{eq:lax6}
\eqlb{eq:LVI}{L^{\mathrm{VI}}=\brc{\matr{cc}{v&0\\0&-v}}+\sum_{\alpha=0}^3L_{\alpha}^{\mathrm{VI}},\qq L_{\alpha}^{\mathrm{VI}}=\brc{\matr{cc}{
0&\nu_{\alpha}\varphi_{\alpha}\brc{u+\omega_{\alpha},z}\\
\nu_{\alpha}\varphi_{\alpha}\brc{-u+\omega_{\alpha},z}&0\\
}},}
\eqlb{eq:MVI}{M^{\mathrm{VI}}=\sum_{\alpha=0}^3 M^{\mathrm VI}_{\alpha},\qq M^{\mathrm VI}_{\alpha}=\brc{\matr{cc}{
0&\nu_{\alpha}f_{\alpha}\brc{u+\omega_{\alpha},z}\\
\nu_{\alpha}f_{\alpha}\brc{-u+\omega_{\alpha},z}&0\\
}},}
\end{subequations}
where functions \eqs{\varphi_{\alpha}, \; f_{\alpha}} \cite{Krichever'1980} are defined in the following way (see Appendix \ref{app:ellipticfunctions}):
\eq{\varphi_{\alpha}\brc{u+\omega_{\beta},z}=\mbe\brc{z\partial_{\tau}\omega_{\alpha}}\phi\brc{u+\omega_{\beta},z},}
\eq{f_{\alpha}\brc{u+\omega_{\beta},z}=
\mbe\brc{z\partial_{\tau}\omega_{\alpha}}\partial_w\phi\brc{w,z}|_{w=u+\omega_{\beta}},}
\eq{\phi(u,z)=\dfrac{\theta_{11}(u+z)\theta_{11}'(0)}{\theta_{11}(u)\theta_{11}(z)}.}

\section{Linear problem for the Painlev\'{e} V equation}
\label{sec:PV}
In order to obtain a relation between linear problems for the Painlev\'{e} equations VI and V we consider two different degeneration procedures. These procedures give linear problems for the following ordinary differential equations:
\eqlb{eq:p5_1}{\dfrac{\rmd^2 u}{\rmd t^2}=C_0^2\dfrac{\cos u}{\sin^3 u}+C_1^2\dfrac{\sin u}{\cos^3 u}+C_2^2\rme^{2t}\sin(4u)+C_2C_3\rme^{t}\sin(2u),}
\eqlb{eq:p5_2}{\dfrac{\rmd^2 u}{\rmd t^2}=C_0^2\dfrac{\cos u}{\sin^3 u}+C_1^2\dfrac{\sin u}{\cos^3 u}+C_4^2\rme^{t}\sin(2u),}
which are the particular cases of the Painlev\'{e} V equation (\ref{eq:PVB}). Equations (\ref{eq:p5_1}) and (\ref{eq:p5_2}) describe the Painlev\'{e} V equation with any choice of arbitrary constants. Even though these equations are connected by the following limit: \eqlb{eq:pcl}{C_2\rightarrow0,\qq C_3\rightarrow\infty,\qq C_2C_3\rightarrow\textrm{const}=C^2_4,}
the Lax pair of the linear problem for equation (\ref{eq:p5_1}) (obtained in Section \ref{sec:p5_1})
diverges upon taking (\ref{eq:pcl}). Thus, we construct linear problems for equations (\ref{eq:p5_1})
and (\ref{eq:p5_2}) separately.

In both degeneration procedures we use the following decomposition of the parameter \eqs{\tau} of an elliptic curve:
\eqlb{eq:spdec}{\tau=\tau_1+\tau_2,}
where \eqs{\tau_1} stands for the time in the limiting system and \eqs{\tau_2} gives the trigonometric limit \eqs{Im\,\tau_2\rightarrow+\infty}. The difference between the degeneration procedures is due to the infinite shifts of coordinate \eqs{u} and spectral parameter \eqs{z}. The degeneration procedures also differ in the scalings of constants of the linear problem for the Painlev\'{e} VI equation.

We will start with the degeneration procedure giving the linear problem for equation (\ref{eq:p5_1}).

\subsection{Linear problem for equation (\ref{eq:p5_1})}
\label{sec:p5_1}
We decompose the parameter  \eqs{\tau} of an elliptic curve as it was described earlier, \eqs{\tau=\tau_1+\tau_2}, which implies
\eq{\dfrac{\rmd u}{\rmd\tau}=\dfrac{\rmd u}{\rmd\tau_1}.}
The scalings of coupling constants are defined by the limiting behavior of Lax matrices (\ref{eq:LVI}), (\ref{eq:MVI}) as follows:
\eq{\nu_0=\dfrac{\wt\nu_0}{\pi},\qquad\nu_1=\dfrac{\wt\nu_1}{\pi}, \qquad\nu_2=\dfrac{-\wt\nu_2q_2^{-\frac12}+\wt\nu_3}{2\pi}, \qquad\nu_3=\dfrac{\wt\nu_2q_2^{-\frac12}+\wt\nu_3}{2\pi},}
where \eqs{q_2\equiv\mbe\brc{\tau_2}.} Thus, we obtain the limiting Hamiltonian and the linear problem of the following form:
\eqlb{eq:HPV}{H^{\mathrm{V}}=v^2-\dfrac{\wt\nu_0^2}{\sin^2(\pi u)}-\dfrac{\wt\nu_1^2}{\cos^2(\pi u)}-8q_1^{\frac12}\wt\nu_2\wt\nu_3\cos(2\pi u)+8q_1\wt\nu_2^2\cos(4\pi u),}
\eqlb{eq:LEPV}{\partial_{\tau_1} L^{\mathrm{V}}-\dfrac1{2\pi\rmi}\partial_z M^{\mathrm{V}}=\bsq{L^{\mathrm{V}},M^{\mathrm{V}}}=\bfi{H^{\mathrm{V}},L^{\mathrm{V}}},}
where
\eq{H^{\mathrm{V}}=\lim_{Im\,\tau_2\rightarrow+\infty}H^{\mathrm{VI}},\qq q_1\equiv\mbe\brc{\tau_1},}
\begin{subequations}
\label{eq:lax5_1}
\eqlb{eq:LV}{L^{\mathrm{V}}=\lim_{Im\,\tau_2\rightarrow+\infty}L^{\mathrm{VI}}=\brc{\matr{cc}{v&0\\0&-v}}+\sum_{\alpha=0}^3\wt\nu_{\alpha}L^{\mathrm{V}}_{\alpha},} \eq{\matr{ll}{
L^{\mathrm{V}}_0=\brc{\matr{cc}{
0&\ctg(\pi u)+\ctg(\pi z)\\
-\ctg(\pi u)+\ctg(\pi z)&0
}},&
L^{\mathrm{V}}_1=\brc{\matr{cc}{
0&\ctg(\pi z)-\tg(\pi u)\\
\ctg(\pi z)+\tg(\pi u)&0\\
}},\svs
L^{\mathrm{V}}_2=4q_1^{\frac12}\brc{\matr{cc}{
0&-\sin\brc{\pi(2u+z)}\\
\sin\brc{\pi(2u-z)}&0
}},&
L^{\mathrm{V}}_3=\dfrac1{\sin(\pi z)}\brc{\matr{cc}{
0&1\\
1&0
}},
}}
\eqlb{eq:MV}{M^{\mathrm{V}}=\lim_{Im\,\tau_2\rightarrow+\infty}M^{\mathrm{VI}}=\sum_{\alpha=0}^3\wt\nu_{\alpha}M^{\mathrm{V}}_{\alpha},}
\eq{\matr{ll}{
M^{\mathrm{V}}_0=-\dfrac{\pi}{\sin^2\brc{\pi u}}\brc{\matr{cc}{
0&1\\
1&0
}},&
M^{\mathrm{V}}_1=-\dfrac{\pi}{\cos^2\brc{\pi u}}\brc{\matr{cc}{
0&1\\
1&0
}},\svs
M^{\mathrm{V}}_2=-8\pi q_1^{\frac12}\brc{\matr{cc}{
0&\cos\brc{\pi(2u+z)}\\
\cos\brc{\pi(2u-z)}&0
}},&
M^{\mathrm{V}}_3=0.
}}
\end{subequations}
We will also need an equivalent form of the Hamiltonian to calculate the limit in Section \ref{V->IV}. This form can be derived from expression (\ref{eq:ham6z}) in the following way:
\eqal{eq:HPVz}{&&H^{\mathrm{V}}\cong H^{\mathrm{V}}-F(z)=v^2-\wt\nu_0^2\brc{\dfrac1{\sin^2(\pi u)}-\dfrac1{\sin^2(\pi z)}}-\wt\nu_1^2\brc{\dfrac1{\cos^2(\pi u)}-\dfrac1{\cos^2(\pi z)}}-\nonumber\\
&&-8q_1^{\frac12}\wt\nu_2\wt\nu_3\brc{\cos(2\pi u)-\cos(2\pi z)}+8q_1\wt\nu_2^2\brc{\cos(4\pi u)-\cos(4\pi z)}.}

Limiting Hamiltonian (\ref{eq:HPV}) coincides with the one known for the fifth Painlev\'{e} equation \cite{Takasaki'01}. It is useful to note that (\ref{eq:LEPV}) is equivalent to (\ref{eq:p5_1}), which is a particular case of the Painlev\'{e} V equation. Indeed, we rewrite (\ref{eq:LEPV}) as a system of two first-order differential equations
\eq{\left\{\matr{l}{
\dfrac{\rmd u}{\rmd\tau_1}=2v,\svs
\dfrac{\rmd v}{\rmd\tau_1}=-2\pi\wt\nu_0^2\dfrac{\cos(\pi u)}{\sin^3(\pi u)}+2\pi\wt\nu_1^2\dfrac{\sin(\pi u)}{\cos^3(\pi u)}-16\pi q_1^{\frac12}\wt\nu_2\wt\nu_3\sin(2\pi u)+32\pi q_1\wt\nu_2^2\sin(4\pi u),
}\right.}
which gives
\eq{\dfrac{\rmd^2 u}{\rmd\tau_1^2}=-4\pi\wt\nu_0^2\dfrac{\cos(\pi u)}{\sin^3(\pi u)}+4\pi\wt\nu_1^2\dfrac{\sin(\pi u)}{\cos^3(\pi u)}-32\pi q_1^{\frac12}\wt\nu_2\wt\nu_3\sin(2\pi u)+64\pi q_1\wt\nu_2^2\sin(4\pi u).}

\subsection{Linear problem for equation (\ref{eq:p5_2})}
\label{sec:p5_2}
In this subsection besides decomposition (\ref{eq:spdec}) of the parameter \eqs{\tau} the following shifts are used:
\eqlb{eq:5uzsub}{u=\wt u-\dfrac\tau2,\qq z=\wt z-\dfrac\tau2.}
The scalings of coupling constants are defined by the limiting behavior of Hamiltonian (\ref{eq:ham6}) as follows:
\eq{\nu_0=\dfrac{\rmi\wt\nu_2 q_2^{-1/4}}{\sqrt2\pi},\qq\nu_1=\dfrac{\wt\nu_3q_2^{-1/4}}{\sqrt2\pi},\qq\nu_2=\dfrac{\wt\nu_0}{\pi},\qq\nu_3=\dfrac{\wt\nu_1}{\pi}.}

Since \eqs{\tau_1} stands for the time in the limiting system, the shifts (\ref{eq:5uzsub}) are time-dependent, hence,
\eqlb{eq:sub65u}{\dfrac{\rmd u}{\rmd\tau}=\dfrac{\rmd u}{\rmd\tau_1}=\dfrac{\rmd\wt u}{\rmd\tau_1}-\dfrac12,}
and the Hamiltonian defining the equations of motion of the limiting system is given by
\eq{H^{\mathrm{V}}=\lim_{Im\,\tau_2\rightarrow+\infty}H^{\mathrm{VI}}+\dfrac12v+\dfrac1{16}.}
Using the Hamiltonian (\ref{eq:ham6}) of the Painlev\'{e} VI equation we get
\eqlb{eq:p5_2ham}{H^{\mathrm{V}}=\brc{v+\dfrac14}^2-\dfrac{\wt\nu_0^2}{\sin^2\brc{\pi\wt u}}-\dfrac{\wt\nu_1^2}{\cos^2\brc{\pi\wt u}}-4\brc{\wt\nu_2^2+\wt\nu_3^2}q_1^{1/2}\cos\brc{2\pi\wt u}.}

In order to obtain convergent Lax matrices it is necessary to perform the gauge transformation of the form
\eq{L^{\mathrm{VI}}\rightarrow gL^{\mathrm{VI}}g^{-1},\qq M^{\mathrm{VI}}\rightarrow gM^{\mathrm{VI}}g^{-1},}
\eq{g=\brc{\matr{cc}{
1&0\\
0&q_2^{-1/4}
}}.}
Since the shift of the spectral parameter in the degeneration procedure under consideration is time-dependent, equation (\ref{eq:idp6}) turns into
\eq{\partial_{\tau_1}L^{\mathrm{VI}}-\partial_{\wt z}\brc{\dfrac1{2\pi\rmi}M^{\mathrm{VI}}-\dfrac12L^{\mathrm{VI}}}=\bsq{L^{\mathrm{VI}},M^{\mathrm{VI}}},}
where \eqs{L^{\mathrm{VI}}=L^{\mathrm{VI}}\brc{\wt u-\tau/2,v,\wt z-\tau/2,\tau},\quad
M^{\mathrm{VI}}=M^{\mathrm{VI}}\brc{\wt u-\tau/2,\wt z-\tau/2,\tau}.}
Thus, the Lax pair of the linear problem for equation (\ref{eq:p5_2}) is defined via
\eq{L^{\mathrm{V}}=2\pi\rmi\lim_{Im \tau_2\rightarrow+\infty}g L^{\mathrm{VI}}g^{-1},\qq M^{\mathrm{V}}=\lim_{Im \tau_2\rightarrow+\infty}g\brc{M^{\mathrm{VI}}-\pi\rmi L^{\mathrm{VI}}}g^{-1}.}
Equation of zero curvature (\ref{eq:idp6}) takes the following form in the limit:
\eqlb{eq:LEPV2}{\partial_{\tau_1}L^{\mathrm{V}}-\partial_{\wt z}M^{\mathrm{V}}=\bsq{L^{\mathrm{V}},M^{\mathrm{V}}},}
where
\eqlb{eq:p5_2L0}{L^{\mathrm{V}}=2\pi\rmi\brc{\matr{cc}{v&0\\0&-v}}+\sum_{\alpha=0}^{3}\wt\nu_{\alpha}L_{\alpha}^{\mathrm{V}},}
\eq{L^{\mathrm{V}}_0=\dfrac{2\pi\rmi}{\sin\brc{\pi\wt u}}\brc{\matr{cc}{
0&\rme^{\rmi\pi\brc{\wt u+\wt z}}q_1^{-1/4}\\
-\rme^{-\rmi\pi\brc{\wt u+\wt z}}q_1^{1/4}&0
}},\qq
L^{\mathrm{V}}_1=\dfrac{2\pi\rmi}{\cos\brc{\pi\wt u}}\brc{\matr{cc}{
0&\rmi\rme^{\rmi\pi\brc{\wt u+\wt z}}q_1^{-1/4}\\
\rmi\rme^{-\rmi\pi\brc{\wt u+\wt z}}q_1^{1/4}&0
}},}
\eq{L^{\mathrm{V}}_2=2\sqrt2\pi\rmi\brc{\matr{cc}{
0&\rme^{2\pi\rmi\brc{\wt u+\wt z}}-1\\
-2\rmi\rme^{-\rmi\pi\brc{\wt u+\wt z}}q_1^{1/2}\sin\brc{\pi\brc{\wt u-\wt z}}&0
}},}
\eq{L^{\mathrm{V}}_3=-2\sqrt2\pi\brc{\matr{cc}{
0&\rme^{2\pi\rmi\brc{\wt u+\wt z}}+1\\
2\rme^{-\rmi\pi\brc{\wt u+\wt z}}q_1^{1/2}\cos\brc{\pi\brc{\wt u-\wt z}}&0
}},}
\eqlb{eq:p5_2M0}{M^{\mathrm{V}}=\rmi\pi\brc{\matr{cc}{
-v&0\\
0&v
}}+\sum_{\alpha=0}^{3}\wt\nu_{\alpha}M^{\mathrm{V}}_{\alpha},}
\eq{M^{\mathrm{V}}_0=-\pi\dfrac{\cos\brc{\pi\wt u}}{\sin^2\brc{\pi\wt u}}\brc{\matr{cc}{
0&q_1^{-1/4}\rme^{\rmi\pi\brc{\wt u+\wt z}}\\
q_1^{1/4}\rme^{-\rmi\pi\brc{\wt u+\wt z}}&0
}},}
\eq{M^{\mathrm{V}}_1=\rmi\pi\dfrac{\sin\brc{\pi\wt u}}{\cos^2\brc{\pi\wt u}}\brc{\matr{cc}{
0&q_1^{-1/4}\rme^{\rmi\pi\brc{\wt u+\wt z}}\\
-q_1^{1/4}\rme^{-\rmi\pi\brc{\wt u+\wt z}}&0
}},}
\eq{M^{\mathrm{V}}_2=\sqrt2\pi\rmi\brc{\matr{cc}{
0&1+\rme^{2\rmi\pi\brc{\wt u+\wt z}}\\
q_1^{1/2}\brc{\rme^{-2\rmi\pi\wt z}+\rme^{-2\rmi\pi\wt u}}&0
}},}
\eq{M^{\mathrm{V}}_3=\sqrt2\pi\brc{\matr{cc}{
0&1-\rme^{2\rmi\pi\brc{\wt u+\wt z}}\\
q_1^{1/2}\brc{\rme^{-2\rmi\pi\wt z}-\rme^{-2\rmi\pi\wt u}}&0
}}.}

Lax pair (\ref{eq:p5_2L0}), (\ref{eq:p5_2M0}) can be simplified by means of the following
gauge transformation:
\eqlb{eq:p5_2gauge}{\wt L^{\mathrm{V}}=\wt gL^{\mathrm{V}}\wt g^{-1}-\brc{\partial_{\wt z}\wt g}\wt g^{-1},\qq
\wt M^{\mathrm{V}}=\wt gM^{\mathrm{V}}\wt g^{-1}-\brc{\partial_{\tau_1}\wt g}\wt g^{-1},}
\eq{\wt g=\brc{\matr{cc}{
q_1^{1/8}\rme^{-\rmi\pi\brc{\wt u+\wt z}/2}&0\\
0&q_1^{-1/8}\rme^{\rmi\pi\brc{\wt u+\wt z}/2}
}}.}
After this transformation the coordinate velocity \eqs{v} enters into the Lax matrix \eqs{\wt L^{\mathrm{V}}}
with the same shift \eqs{v+\frac14} as in the Hamiltonian (\ref{eq:p5_2ham}), i.e.,
\begin{subequations}
\label{eq:lax5_2}
\eqlb{eq:P5_2Lg}{\wt L^{\mathrm{V}}=2\pi\rmi\brc{\matr{cc}{
v+\dfrac14&0\\
0&-v-\dfrac14
}}+\sum_{\alpha=0}^3\wt\nu_{\alpha}\wt L^{\mathrm{V}}_{\alpha},}
\eq{\wt L^{\mathrm{V}}_0=\dfrac{2\pi\rmi}{\sin\brc{\pi\wt u}}\brc{\matr{cc}{
0&1\\
-1&0
}},\qq
\wt L^{\mathrm{V}}_1=-\dfrac{2\pi}{\cos\brc{\pi\wt u}}\brc{\matr{cc}{
0&1\\
1&0
}},}
\eq{\wt L^{\mathrm{V}}_2=4\sqrt2\pi q_1^{1/4}\brc{\matr{cc}{
0&-\sin\brc{\pi\brc{\wt u+\wt z}}\\
\sin\brc{\pi\brc{\wt u-\wt z}}&0
}},}
\eq{\wt L^{\mathrm{V}}_3=-4\sqrt2\pi q_1^{1/4}\brc{\matr{cc}{
0&\cos\brc{\pi\brc{\wt u+\wt z}}\\
\cos\brc{\pi\brc{\wt u-\wt z}}&0
}}.}

Using the Hamilton equation of motion for the coordinate \eqs{\wt u}
\eq{\dfrac{\rmd\wt u}{\rmd\tau_1}=\bfi{H^{\mathrm{V}},\wt u}=2v+\dfrac12,}
one can ensure that transformation (\ref{eq:p5_2gauge}) removes \eqs{v}
from the second Lax matrix (\ref{eq:p5_2M0}), namely,
\eqlb{eq:p5_2Mg}{\wt M^{\mathrm{V}}=\sum_{\alpha=0}^3\wt\nu_{\alpha}\wt M^{\mathrm{V}}_{\alpha},}
\eq{\wt M^{\mathrm{V}}_0=-\pi\dfrac{\cos\brc{\pi\wt u}}{\sin^2\brc{\pi\wt u}}\brc{\matr{cc}{
0&1\\
1&0
}},\qq
\wt M^{\mathrm{V}}_1=\rmi\pi\dfrac{\sin\brc{\pi\wt u}}{\cos^2\brc{\pi\wt u}}\brc{\matr{cc}{
0&1\\
-1&0
}},}
\eq{\wt M^{\mathrm{V}}_2=2\sqrt2\pi\rmi q_1^{1/4}\brc{\matr{cc}{
0&\cos\brc{\pi\brc{\wt u+\wt z}}\\
\cos\brc{\pi\brc{\wt u-\wt z}}&0
}},}
\eq{\wt M^{\mathrm{V}}_3=2\sqrt2\pi\rmi q_1^{1/4}\brc{\matr{cc}{
0&-\sin\brc{\pi\brc{\wt u+\wt z}}\\
\sin\brc{\pi\brc{\wt u-\wt z}}&0
}}.}
\end{subequations}

To show that equations (\ref{eq:LEPV2}) and (\ref{eq:p5_2}) are equivalent we rewrite (\ref{eq:LEPV2})
in the form of a system of two first-order differential equations
\eq{\left\{\matr{l}{
\dfrac{\rmd\wt u}{\rmd\tau_1}=2v+\dfrac12,\svs
\dfrac{\rmd v}{\rmd\tau_1}=-2\pi\wt\nu_0^2\dfrac{\cos(\pi\wt u)}{\sin^3(\pi\wt u)}+2\pi\wt\nu_1^2\dfrac{\sin(\pi\wt u)}{\cos^3(\pi\wt u)}-8\pi q_1^{\frac12}\brc{\wt\nu_2^2+\wt\nu_3^2}\sin(2\pi\wt u).
}\right.}
Eliminating \eqs{v} we obtain
\eq{\dfrac{\rmd^2\wt u}{\rmd\tau_1^2}=-4\pi\wt\nu_0^2\dfrac{\cos(\pi\wt u)}{\sin^3(\pi\wt u)}+4\pi\wt\nu_1^2\dfrac{\sin(\pi\wt u)}{\cos^3(\pi\wt u)}-16\pi q_1^{\frac12}\brc{\wt\nu_2^2+\wt\nu_3^2}\sin(2\pi\wt u).}

\section{Linear problem for the Painlev\'{e} IV equation}
\label{V->IV}

We construct a linear problem for the Painlev\'{e} IV equation as the limit of the linear problem
for the Painlev\'{e} V equation obtained in Section \ref{sec:p5_1}. We make the substitutions
\eqlb{eq:subIV}{\tau_1=\dfrac{t w^2}{2\pi\rmi},\qquad u=\wt u\dfrac w{2\pi\rmi},\qquad z=\wt z\dfrac w{2\pi\rmi},\qquad v=\dfrac{\wt v}w,}
the scalings of coupling constants and the limit itself \eqs{w\rightarrow0}. After applying (\ref{eq:subIV})
the canonical Poisson bracket acquires the following form:
\eq{\bfi{\wt v,\wt u}=2\pi\rmi.}

We define scalings of coupling constants by the limiting behavior of equations of motion (\ref{eq:LEPV}) and the Lax matrices (\ref{eq:LV}), (\ref{eq:MV}) via
\eq{\wt\nu_0=\dfrac{\beta}{4\sqrt{2}},\qquad\wt\nu_1=-2\rmi w^{-4},\qquad\wt\nu_2=\dfrac{\rmi}{4w^4},\qquad\wt\nu_3=\dfrac{8\rmi+w^4\rmi\alpha}{4w^4}.}
To obtain the Hamiltonian of the limiting system we use Hamiltonian (\ref{eq:HPVz}) for the
Painlev\'{e} V equation, because the other Hamiltonian (\ref{eq:HPV}) diverges as
\eqs{w\rightarrow0}. Thus, the Hamiltonian and the Lax matrices of the limiting system are
defined as
\eq{H^{\textrm{IV}}=\lim_{w\rightarrow0}\dfrac{w^2}{2\pi\rmi}H^{\mathrm{V}},\qq L^{\mathrm{IV}}=\lim_{w\rightarrow0}w L^{\mathrm{V}},\qq M^{\mathrm{IV}}=\lim_{w\rightarrow0}\dfrac{w^2}{2\pi\rmi}M^{\mathrm{V}}.}

Finally, we get the following limiting Hamiltonian and the equation of zero curvature:
\eq{H^{\mathrm{IV}}=-\dfrac{\rmi\wt v^2}{2\pi}+\dfrac{\rmi\brc{\wt u^6-\wt z^6}}{32\pi}+\dfrac{\rmi t\brc{\wt u^4-\wt z^4}}{8\pi}+\dfrac{\rmi\brc{t^2-\alpha}\brc{\wt u^2-\wt z^2}}{8\pi}-\dfrac{\rmi\beta^2}{16\pi}\brc{\dfrac1{\wt u^2}-\dfrac1{\wt z^2}},}
\eqlb{eq:linIV}{\partial_tL^{\mathrm{IV}}-\partial_{\wt z}M^{\mathrm{IV}}=\bsq{L^{\mathrm{IV}},M^{\mathrm{IV}}}=\bfi{H^{\mathrm{IV}},L^{\mathrm{IV}}},}
where
\normalsize
\begin{subequations}
\label{eq:lax4}
\eqlb{eq:p_4L}{L^{\mathrm{IV}}=\brc{\matr{cc}{
\wt v&-\dfrac{\wt u^3}4-\dfrac{\wt u^2\wt z}4-\wt u\dfrac{\wt z^2+4t}8-\dfrac{\wt z^3}{32}-\dfrac{t\wt z}4\\
\dfrac{\wt u^3}4-\dfrac{\wt u^2\wt z}4+\wt u\dfrac{\wt z^2+4t}8-\dfrac{\wt z^3}{32}-\dfrac{t\wt z}4&-\wt v
}}+}
\eq{+\brc{\matr{cc}{
0&-\dfrac{\alpha}{2\wt z}+\dfrac{\rmi\beta}{2\sqrt2}\brc{\dfrac1{\wt u}+\dfrac1{\wt z}}\\
-\dfrac{\alpha}{2\wt z}+\dfrac{\rmi\beta}{2\sqrt2}\brc{\dfrac1{\wt z}-\dfrac1{\wt u}}&0

}},}
\eqlb{eq:p_4M}{M^{\mathrm{IV}}=\brc{\matr{cc}{
0&-\dfrac{3\wt u^2}4-\dfrac{\wt u\wt z}2-\dfrac t2-\dfrac{\wt z^2}8-\dfrac{\rmi\beta}{2\sqrt2\wt u^2}\\
-\dfrac{3\wt u^2}4+\dfrac{\wt u\wt z}2-\dfrac t2-\dfrac{\wt z^2}8-\dfrac{\rmi\beta}{2\sqrt2\wt u^2}&0
}}.}
\end{subequations}

The equivalence of equation (\ref{eq:linIV}) to the Painlev\'{e} IV equation in the form (\ref{eq:PIVB})
can be shown in two steps. First, we rewrite (\ref{eq:linIV}) as a system of two differential equations
\eq{\left\{\matr{l}{
\dfrac{\rmd\wt u}{\rmd t}=2\wt v,\svs
\dfrac{\rmd\wt v}{\rmd t}=\dfrac{\beta^2}{4\wt u^3}+\dfrac12\brc{t^2-\alpha}\wt u+t\wt u^3+\dfrac{3\wt u^5}8.
}\right.}
Second, after eliminating \eqs{v} from the system we get the following second-order differential equation:
\eq{\dfrac{\rmd^2\wt u}{\rmd t^2}=\dfrac{\beta^2}{2\wt u^3}+\brc{t^2-\alpha}\wt u+2t\wt u^3+\dfrac{3\wt u^5}{4}.}

\section{Linear problem for the Painlev\'e III equation}
\label{sec:PIII}
As in Section \ref{sec:PV} we construct a limiting procedure which transforms the linear problem (\ref{eq:lax6}) for the Painlev\'e VI equation into linear problems for the following two equations:
\eqlb{eq:p3_1}{\dfrac{\rmd^2 u}{\rmd t^2}=C_0^2\rme^{2t+2u}+C_2^2\rme^{2t-2u}+C_0C_1\rme^{t+u}+C_2C_3\rme^{t-u},}
\eqlb{eq:p3_2}{\dfrac{\rmd^2 u}{\rmd t^2}=C_2^2\rme^{2t-2u}+C_0^2\rme^{t+u}+C_2C_3\rme^{t-u}.}
Equations (\ref{eq:p3_1}) and (\ref{eq:p3_2}) describe the Painlev\'e III equation (\ref{eq:PIIIB}) with any choice of arbitrary constants. In Sections \ref{sec:4.1} and \ref{sec:4.2} we construct two distinct degeneration procedures which give different linear problems for equation (\ref{eq:p3_1}). A linear problem for equation (\ref{eq:p3_2}) is constructed in Section \ref{sec:4.3}.

Degeneration procedures under consideration are based on a generalization of the Inozemtsev limit and differ in shifts of the spectral parameter and scalings of coupling constants. The generalization of the Inozemtsev limit consists of the decomposition of the parameter \eqs{\tau}
\eqlb{eq:3tdec}{\tau=\tau_1+\tau_2,}
with \eqs{\tau_1} denoting the time of the system,
the shift of the coordinate
\eqlb{eq:3usub}{u=\wt u + \dfrac\tau4,}
and the trigonometric limit \eqs{Im\,\tau_2\rightarrow+\infty}.

\subsection{First linear problem for equation (\ref{eq:p3_1})}
\label{sec:4.1}
To construct a linear problem associated with equation (\ref{eq:p3_1}) we use decomposition (\ref{eq:3tdec}), shift of the coordinate \eqs{u} (\ref{eq:3usub}), and the trigonometric limit \eqs{Im\,\tau_2\rightarrow+\infty}. From the decomposition of the Lax matrices (\ref{eq:LVI}), (\ref{eq:MVI}) as a series in \eqs{q} one can determine the scalings of coupling constants
\eqlb{eq:sub53}{\nu_0=\dfrac{\wt\nu_0q_2^{-\frac14}+\wt\nu_1}{2\pi},\qq\nu_1=\dfrac{-\wt\nu_0q_2^{-\frac14}+\wt\nu_1}{2\pi},\qq \nu_2=\dfrac{\wt\nu_2q_2^{-\frac14}+\wt\nu_3}{2\pi},\qq\nu_3=\dfrac{-\wt\nu_2q_2^{-\frac14}+\wt\nu_3}{2\pi}.}

Since \eqs{\tau_1} is the time of the system, shift (\ref{eq:3usub}) of the coordinate \eqs{u} is time-dependent, which implies
\eqlb{eq:sub53u}{\dfrac{\rmd u}{\rmd\tau}=\dfrac{\rmd u}{\rmd\tau_1}=\dfrac{\rmd\wt u}{\rmd\tau_1}+\dfrac14.}
Thus, the Hamiltonian of the limiting system has the following form:
\eq{H^{\mathrm{III}}=\lim_{Im\,\tau_2\rightarrow+\infty}H^{\mathrm{VI}}-\dfrac14v+\dfrac1{64}.}
Using the Hamiltonian for the Painlev\'e VI equation in the form (\ref{eq:ham6}), we get
\eqlb{eq:ham3}{H^{\mathrm{III}}=\brc{v-\dfrac18}^2+4q_1^{\frac12}\wt\nu_0^2\rme^{4\pi\rmi\wt u}+4q_1^{\frac12}\wt\nu_2^2\rme^{-4\pi\rmi\wt u}+4q_1^{\frac14}\wt\nu_0\wt\nu_1\rme^{2\pi\rmi\wt u}+4q_1^{\frac14}\wt\nu_2\wt\nu_3\rme^{-2\pi\rmi\wt u}.}
Expression (\ref{eq:ham6z}) gives an equivalent Hamiltonian
\eqal{eq:ham3z}{&H^{\mathrm{III}}\cong H^{\mathrm{III}}-F(z)=\brc{v-\dfrac18}^2+4q_1^{\frac12}\wt\nu_0^2\brc{\rme^{4\pi\rmi\wt u}-\rme^{4\pi\rmi z}}+4q_1^{\frac12}\wt\nu_2^2\brc{\rme^{-4\pi\rmi\wt u}-\rme^{-4\pi\rmi z}}+\nonumber\\
&+4q_1^{\frac14}\wt\nu_0\wt\nu_1\brc{\rme^{2\pi\rmi\wt u}-\rme^{2\pi\rmi z}}+4q_1^{\frac14}\wt\nu_2\wt\nu_3\brc{\rme^{-2\pi\rmi\wt u}-\rme^{-2\pi\rmi z}},}
which will be used to calculate a limit in Section \ref{sec:PII}.

The equation of zero curvature (\ref{eq:idp6}) preserves the form in the limit
\eqlb{eq:mo3}{\partial_{\tau_1}L^{\mathrm{III}}-\dfrac1{2\pi\rmi}\partial_z M^{\mathrm{III}}=\bsq{L^{\mathrm{III}},M^{\mathrm{III}}},}
where
\begin{subequations}
\label{eq:lax3_1}
\eqlb{eq:L3}{L^{\mathrm{III}}=\lim_{Im\,\tau_2\rightarrow+\infty}L^{\mathrm{VI}}=\brc{\matr{cc}{v&0\\0&-v}}+\sum_{\alpha=0}^3\wt\nu_{\alpha}L^{\mathrm{III}}_{\alpha},}

\eq{\matr{ll}{
L^{\mathrm{III}}_0=2\rmi q_1^{\frac14}\rme^{2\pi\rmi\wt u}\brc{\matr{cc}{0&-1\\1&0}},&
L^{\mathrm{III}}_1=\brc{\matr{cc}{
0&-\rmi+\ctg\brc{\pi z}\\
\rmi+\ctg\brc{\pi z}&0
}},\svs
L^{\mathrm{III}}_2=2\rmi q_1^{\frac14}\brc{\matr{cc}{
0&\rme^{-\rmi\pi\brc{2\wt u+z}}\\
-\rme^{-\rmi\pi\brc{2\wt u-z}}&0
}},&
L^{\mathrm{III}}_3=\dfrac1{\sin\brc{\pi z}}\brc{\matr{cc}{0&1\\1&0}},
}}

\eqlb{eq:M3}{M^{\mathrm{III}}=\lim_{Im\,\tau_2\rightarrow+\infty}L^{\mathrm{VI}}=\sum_{\alpha=0}^3\wt\nu_{\alpha}M^{\mathrm{III}}_{\alpha},}

\eq{\matr{ll}{M^{\mathrm{III}}_0=4\pi q_1^{\frac14}\rme^{2\pi\rmi\wt u}\brc{\matr{cc}{0&1\\1&0}},\qq&
M^{\mathrm{III}}_1=0,\svs
M^{\mathrm{III}}_2=4\pi q_1^{\frac14}\brc{\matr{cc}{
0&\rme^{-\rmi\pi\brc{2\wt u+z}}\\
\rme^{-\rmi\pi\brc{2\wt u-z}}&0
}},\qq&
M^{\mathrm{III}}_3=0.
}}
\end{subequations}
Equation (\ref{eq:mo3}) describes the Hamilton equations of motion of the limiting system
\eq{\left\{\matr{l}{
\dfrac{\rmd\wt u}{\rmd\tau_1}=\bfi{H^{\mathrm{III}},\wt u}=2v-\dfrac14,\svs
\dfrac{\rmd v}{\rmd\tau_1}=\bfi{H^{\mathrm{III}},v}=-16\rmi\pi q_1^{\frac12}\wt\nu_0^2\rme^{4\pi\rmi\wt u}+16\rmi\pi q_1^{\frac12}\wt\nu_2^2\rme^{-4\pi\rmi\wt u}-8\rmi\pi q_1^{\frac14}\wt\nu_0\wt\nu_1\rme^{2\pi\rmi\wt u}+8\rmi\pi q_1^{\frac14}\wt\nu_2\wt\nu_3\rme^{-2\pi\rmi\wt u}
}\right.}
and is equivalent to the following second-order differential equation
\eqlb{eq:ep3_1}{\dfrac{\rmd^2 u}{\rmd\tau_1^2}=-32\rmi\pi q_1^{\frac12}\wt\nu_0^2\rme^{4\pi\rmi\wt u}+32\rmi\pi q_1^{\frac12}\wt\nu_2^2\rme^{-4\pi\rmi\wt u}-16\rmi\pi q_1^{\frac14}\wt\nu_0\wt\nu_1\rme^{2\pi\rmi\wt u}+16\rmi\pi q_1^{\frac14}\wt\nu_2\wt\nu_3\rme^{-2\pi\rmi\wt u},}
which in turn coincides with (\ref{eq:p3_1}) up to a change of arbitrary constants.

\subsection{Second linear problem for equation (\ref{eq:p3_1})}
\label{sec:4.2}
Adding to the degeneration procedure described in Section \ref{sec:4.1} the shift of the spectral parameter
\eqlb{eq:p4.1z}{z=\wt z + \dfrac\tau2,}
we get another linear problem associated with equation (\ref{eq:p3_1}). The gauge equivalence of the second linear problem to the first linear problem \ref{eq:lax3_1} has not been established.

Thus, we use decomposition (\ref{eq:3tdec}) of the parameter \eqs{\tau}, shifts of the coordinate (\ref{eq:3usub}) and the spectral parameter (\ref{eq:p4.1z}), and the trigonometric limit \eqs{Im\,\tau_2\rightarrow+\infty}. Scalings of coupling constants are determined from the decomposition of the Hamiltonian (\ref{eq:ham6}) as a series in \eqs{q}
\eq{\nu_0=\dfrac{\wt\nu_0q_2^{-\frac14}+\wt\nu_1}{2\pi},\qq\nu_1=\dfrac{-\wt\nu_0q_2^{-\frac14}+\wt\nu_1}{2\pi},\qq \nu_2=\dfrac{\wt\nu_2q_2^{-\frac14}+\wt\nu_3}{2\pi},\qq\nu_3=\dfrac{\wt\nu_2q_2^{-\frac14}-\wt\nu_3}{2\pi}.}

Since substitution (\ref{eq:3usub}) is time-dependent, the time derivative of the coordinate $u$ of the Calogero-Inozemtsev system and the time derivative of the coordinate \eqs{\wt u} of the limiting system are connected via (\ref{eq:sub53u}) as follows:
\eq{\dfrac{\rmd u}{\rmd\tau}=\dfrac{\rmd u}{\rmd\tau_1}=\dfrac{\rmd\wt u}{\rmd\tau_1}+\dfrac14.}
Thus, the Hamiltonian of the limiting system is of the form
\eq{H^{\mathrm{III}}=\lim_{Im\,\tau_2\rightarrow+\infty}H^{\mathrm{VI}}-\dfrac14v+\dfrac1{64}.}
Using the Hamiltonian (\ref{eq:ham6}) associated with the Painlev\'e VI equation one can derive the following explicit formula for \eqs{H^{\mathrm{III}}}:
\eqlb{eq:ham3_1}{H^{\mathrm{III}}=\brc{v-\dfrac18}^2+4\wt\nu_0^2q_1^{1/2}\rme^{4\pi\rmi\wt u}+4\wt\nu_0\wt\nu_1q_1^{1/4}\rme^{2\pi\rmi\wt u}+4\wt\nu_2^2q_1^{1/2}\rme^{-4\pi\rmi\wt u}+4\wt\nu_2\wt\nu_3q_1^{1/4}\rme^{-2\pi\rmi\wt u}.}

After substitutions (\ref{eq:3tdec}), (\ref{eq:3usub}), and (\ref{eq:p4.1z}) the equation of zero curvature (\ref{eq:idp6}) transforms into
\eq{\partial_{\tau_1}L^{\mathrm{VI}}-\partial_{\wt z}\brc{\dfrac1{2\pi\rmi}M^{\mathrm{VI}}+\dfrac12L^{\mathrm{VI}}}=\bsq{L^{\mathrm{VI}},M^{\mathrm{VI}}},}
where \eqs{L^{\mathrm{VI}}=L^{\mathrm{VI}}\brc{\wt u+\tau/4,v,\wt z+\tau/2,\tau},} and
\eqs{M^{\mathrm{VI}}=M^{\mathrm{VI}}\brc{\wt u+\tau/4,\wt z+\tau/2,\tau}.}
This implies the following definitions of the Lax matrices:
\eq{L^{\mathrm{III}}=2\pi\rmi\lim_{Im \tau_2\rightarrow+\infty} L^{\mathrm{VI}},\qq M^{\mathrm{III}}=\lim_{Im \tau_2\rightarrow+\infty}\brc{M^{\mathrm{VI}}+\pi\rmi L^{\mathrm{VI}}}.}
Finally, the equation of zero curvature acquires the form
\eqlb{eq:p3_1id}{\partial_{\tau_1}L^{\mathrm{III}}-\partial_{\wt z}M^{\mathrm{III}}=\bsq{L^{\mathrm{III}},M^{\mathrm{III}}},}
where
\eqlb{eq:lax3_2}{L^{\mathrm{III}}=\lim_{Im\,\tau_2\rightarrow+\infty}L^{\mathrm{VI}}=2\pi\rmi\brc{\matr{cc}{v&0\\0&-v}}
+\sum_{\alpha=0}^3\wt\nu_{\alpha} L^{\mathrm{III}}_{\alpha},}

\eq{\matr{ll}{
L^{\mathrm{III}}_0=4\pi q_1^{\frac14}\brc{\matr{cc}{
0&\rme^{2\pi\rmi\wt u}-\rme^{-2\pi\rmi\brc{\wt u+\wt z}}\\
-\rme^{2\pi\rmi\wt u}&0
}},&
L^{\mathrm{III}}_1=4\pi\brc{\matr{cc}{0&1\\0&0}},\svs
L^{\mathrm{III}}_2=4\pi q_1^{\frac14}\brc{\matr{cc}{
0&\rme^{\pi\rmi\wt z}-\rme^{-\pi\rmi\brc{4\wt u+\wt z}}\\
\rme^{\pi\rmi\wt z}&0
}},&
L^{\mathrm{III}}_3=-4\pi\rme^{-\pi\rmi\brc{2\wt u+\wt z}}\brc{\matr{cc}{0&1\\0&0}},
}}

\eqlb{eq:laxM3_2}{
M^{\mathrm{III}}=\lim_{Im\,\tau_2\rightarrow+\infty}M^{\mathrm{VI}}=\rmi\pi\brc{\matr{cc}{v&0\\0&-v}}
+\sum_{\alpha=0}^3\wt\nu_{\alpha}
M^{\mathrm{III}}_{\alpha},}
\eq{M^{\mathrm{III}}_0=2\pi q_1^{1/4}\brc{\matr{cc}{
0&\rme^{-2\rmi\pi\brc{\wt u+\wt z}}+3\rme^{2\rmi\pi\wt u}\\
\rme^{2\rmi\pi\wt u}&0
}},\qq
M^{\mathrm{III}}_1=2\pi\brc{\matr{cc}{
0&1\\
0&0
}},}
\eq{M^{\mathrm{III}}_2=2\pi q_1^{1/4}\brc{\matr{cc}{
0&3\rme^{-\rmi\pi\brc{4\wt u+\wt z}}+\rme^{\rmi\pi\wt z}\\
\rme^{\rmi\pi\wt z}&0
}},\qq
M^{\mathrm{III}}_3=2\pi\rme^{-\rmi\pi\brc{2\wt u+\wt z}}\brc{\matr{cc}{0&1\\0&0}}.
}

As in subsection \ref{sec:p5_2} we can remove \eqs{v} from the second Lax matrix \eqs{M^{\mathrm{III}}} (\ref{eq:laxM3_2}) by means of the gauge transformation
\eq{L^{\mathrm{III}}\rightarrow gL^{\mathrm{III}}g^{-1}-\brc{\partial_{\wt z}g}g^{-1},\qq
M^{\mathrm{III}}\rightarrow gM^{\mathrm{III}}g^{-1}-\brc{\partial_{\tau_1}g}g^{-1},}
\eq{g=\brc{\matr{cc}{
q_1^{1/16}\rme^{\rmi\pi\brc{2\wt u+\wt z}/4}&0\\
0&q_1^{-1/16}\rme^{-\rmi\pi\brc{2\wt u+\wt z}/4}
}}.}
Applying the transformation, we get the first Lax matrix \eqs{L^{\mathrm{III}}} (\ref{eq:lax3_2}) with the same shifted velocity \eqs{v-\frac18} as in the Hamiltonian (\ref{eq:ham3_1}).

Equation (\ref{eq:p3_1id}) is equivalent to the Hamilton equations of motion
\eq{\left\{\matr{l}{
\dfrac{\rmd\wt u}{\rmd\tau_1}=\bfi{H^{\mathrm{III}},\wt u}=2v-\dfrac14,\svs
\dfrac{\rmd v}{\rmd\tau_1}=\bfi{H^{\mathrm{III}},v}=-16\pi\rmi\wt\nu_0^2q_1^{1/2}\rme^{4\pi\rmi\wt u}-8\pi\rmi\wt\nu_0\wt\nu_1q_1^{1/4}\rme^{2\pi\rmi\wt u}+16\pi\rmi\wt\nu_2^2q_1^{1/2}\rme^{-4\pi\rmi\wt u}+8\pi\rmi\wt\nu_2\wt\nu_3q_1^{1/4}\rme^{-2\pi\rmi\wt u}.
}\right.}
These equations in turn are equivalent to the following second-order differential equation:
\eq{\dfrac{\rmd^2\wt u}{\rmd\tau_1^2}=-32\pi\rmi\wt\nu_0^2q_1^{1/2}\rme^{4\pi\rmi\wt u}-16\pi\rmi\wt\nu_0\wt\nu_1q_1^{1/4}\rme^{2\pi\rmi\wt u}+32\pi\rmi\wt\nu_2^2q_1^{1/2}\rme^{-4\pi\rmi\wt u}+16\pi\rmi\wt\nu_2\wt\nu_3q_1^{1/4}\rme^{-2\pi\rmi\wt u},}
which coincides with (\ref{eq:p3_1}) up to a change of arbitrary constants.

\subsection{Linear problem for equation (\ref{eq:p3_2})}
\label{sec:4.3}
In this subsection we use decomposition (\ref{eq:3tdec}) of the parameter \eqs{\tau} of an elliptic curve, the substitution of the coordinate (\ref{eq:3usub}), and the shift of the spectral parameter
\eq{z=\wt z +\dfrac{\tau}4.}
The scalings of coupling constants are determined from the decomposition of the Hamiltonian (\ref{eq:ham6}) as a series in \eqs{q} in the following way:
\eq{\nu_0=\dfrac{\wt\nu_0 q_2^{-1/8}}{2\pi},\qq
\nu_1=\dfrac{\rmi\wt\nu_1 q_2^{-1/8}}{2\pi},\qq
\nu_2=\dfrac{\wt\nu_2 q_2^{-1/4}+\wt\nu_3}{2\pi},\qq
\nu_3=\dfrac{-\wt\nu_2 q_2^{-1/4}+\wt\nu_3}{2\pi}.}

As it was mentioned in subsection \ref{sec:4.1} shift of the coordinate (\ref{eq:3usub}) is time-dependent. Thus, the Hamiltonian of the limiting system have the following form
\eq{H^{\mathrm{III}}=\lim_{Im\,\tau_2\rightarrow+\infty}H^{\mathrm{VI}}-\dfrac14v+\dfrac1{64}.}
Using the Hamiltonian (\ref{eq:ham6}) for the Painlev\'e VI equation one can derive
\eqlb{eq:ham3_3}{H^{\mathrm{III}}=\brc{v-\dfrac18}^2+\brc{\wt\nu_0^2+\wt\nu_1^2}q_1^{1/4}\rme^{2\rmi\pi\wt u}+4\wt\nu_2^2q_1^{1/2}\rme^{-4\rmi\pi\wt u}+4\wt\nu_2\wt\nu_3q_1^{1/4}\rme^{-2\rmi\pi\wt u}.}

In this case in order to get convergent Lax matrices it is necessary to make the gauge transformation
\eq{L^{\mathrm{VI}}\rightarrow gL^{\mathrm{VI}}g^{-1},\qq M^{\mathrm{VI}}\rightarrow gM^{\mathrm{VI}}g^{-1},}
\eq{g=\brc{\matr{cc}{
1&0\\
0&q_2^{-1/8}
}}.}
After applying the shift of the spectral parameter, equation of zero curvature (\ref{eq:idp6}) becomes
\eq{\partial_{\tau_1}L^{\mathrm{VI}}-\partial_{\wt z}\brc{\dfrac1{2\pi\rmi}M^{\mathrm{VI}}+\dfrac14L^{\mathrm{VI}}}=\bsq{L^{\mathrm{VI}},M^{\mathrm{VI}}},}
where \eqs{L^{\mathrm{VI}}=L^{\mathrm{VI}}\brc{\wt u+\tau/4,v,\wt z+\tau/4,\tau},\quad
M^{\mathrm{VI}}=M^{\mathrm{VI}}\brc{\wt u+\tau/4,\wt z+\tau/4,\tau}.}
This implies the following definition of the Lax matrices
\eq{L^{\mathrm{III}}=2\pi\rmi\lim_{Im \tau_2\rightarrow+\infty}g L^{\mathrm{VI}}g^{-1},\qq M^{\mathrm{III}}=\lim_{Im \tau_2\rightarrow+\infty}g\brc{M^{\mathrm{VI}}+\dfrac{\pi\rmi}2 L^{\mathrm{VI}}}g^{-1}.}
Finally, the equation of zero curvature acquires the following form:
\eqlb{eq:mo3p}{\partial_{\tau_1}L^{\mathrm{III}}-\partial_{\wt z}M^{\mathrm{III}}=\bsq{L^{\mathrm{III}},M^{\mathrm{III}}},}
where
\begin{subequations}
\label{eq:lax3_3}
\eqlb{eq:p_3L3}{L^{\mathrm{III}}=2\pi\rmi\brc{\matr{cc}{v&0\\0&-v}}+\sum_{\alpha=0}^{3}\wt\nu_{\alpha}L_{\alpha}^{\mathrm{III}},}
\eq{L^{\mathrm{III}}_0=2\pi\brc{\matr{cc}{
0&1\\
q_1^{1/4}\brc{\rme^{2\rmi\pi\wt z}-\rme^{2\rmi\pi\wt u}}&0
}},\qq
L^{\mathrm{III}}_1=2\pi\rmi\brc{\matr{cc}{
0&1\\
q_1^{1/4}\brc{\rme^{2\rmi\pi\wt u}+\rme^{2\rmi\pi\wt z}}&0
}},}
\eq{L^{\mathrm{III}}_2=-4\pi q_1^{1/8}\rme^{-2\rmi\pi\wt u}\brc{\matr{cc}{
0&\rme^{-\rmi\pi\wt z}\\
-q_1^{1/4}\rme^{\rmi\pi\wt z}&0
}},\qq
L^{\mathrm{III}}_3=4\pi q_1^{1/8}\rme^{\rmi\pi\wt z}\brc{\matr{cc}{
0&0\\
1&0
}},}
\eqlb{eq:laxm3_3}{M^{\mathrm{III}}=\dfrac{\rmi\pi}2\brc{\matr{cc}{
v&0\\
0&-v
}}+\sum_{\alpha=0}^{3}\wt\nu_{\alpha}M^{\mathrm{III}}_{\alpha},}
\eq{M^{\mathrm{III}}_0=\dfrac{\pi}2\brc{\matr{cc}{
0&1\\
q_1^{1/4}\brc{3\rme^{2\rmi\pi\wt u}+\rme^{2\rmi\pi\wt z}}&0
}},\qq
M^{\mathrm{III}}_1=\dfrac{\rmi\pi}2\brc{\matr{cc}{
0&1\\
\rme^{2\rmi\pi\wt z}-3\rme^{2\rmi\pi\wt u}&0
}},}
\eq{M^{\mathrm{III}}_2=\pi q_1^{1/8}\brc{\matr{cc}{
0&3\rme^{-\rmi\pi\brc{2\wt u+\wt z}}\\
5q_1^{1/4}\rme^{-\rmi\pi\brc{2\wt u-\wt z}}&0
}},\qq
M^{\mathrm{III}}_3=\pi q_1^{1/8}\rme^{\rmi\pi\wt z}\brc{\matr{cc}{
0&0\\
1&0
}}.}
\end{subequations}

As in subsections \ref{sec:p5_2} and \ref{sec:4.2} we can remove \eqs{v} from the second Lax matrix \eqs{M^{\mathrm{III}}} (\ref{eq:laxm3_3}) by means of the gauge transformation
\eq{L^{\mathrm{III}}\rightarrow\wt gL^{\mathrm{III}}\wt g^{-1}-\brc{\partial_{\wt z}\wt g}\wt g^{-1},\qq
M^{\mathrm{III}}\rightarrow\wt gM^{\mathrm{III}}\wt g^{-1}-\brc{\partial_{\tau_1}\wt g}\wt g^{-1},}
\eq{\wt g=\brc{\matr{cc}{
q_1^{1/32}\rme^{\rmi\pi\brc{\wt u+\wt z}/4}&0\\
0&q_1^{-1/32}\rme^{-\rmi\pi\brc{\wt u+\wt z}/4}
}}.}
Applying this transformation we obtain the first Lax matrix \eqs{L^{\mathrm{III}}} (\ref{eq:p_3L3}) with the same shifted velocity \eqs{v-\frac18} as in the Hamiltonian (\ref{eq:ham3_3}).

Equation of zero curvature (\ref{eq:mo3p}) is equivalent to the Hamilton equations of motion
\eq{\left\{\matr{l}{
\dfrac{\rmd\wt u}{\rmd\tau_1}=\bfi{H^{\mathrm{III}},\wt u}=2v-\dfrac14,\svs
\dfrac{\rmd v}{\rmd\tau_1}=-2\rmi\pi q_1^{1/4}\brc{\wt\nu_0^2+\wt\nu_1^2}\rme^{2\rmi\pi\wt u}+16\rmi\pi q_1^{1/2}\wt\nu_2^2\rme^{-4\rmi\pi\wt u}+8\rmi\pi q_1^{1/4}\wt\nu_2\wt\nu_3\rme^{-2\rmi\pi\wt u}.
}\right.}
Eliminating \eqs{v} from this system we get the second-order differential equation
\eq{\dfrac{\rmd^2\wt u}{\rmd\tau_1^2}=-4\rmi\pi q_1^{1/4}\brc{\wt\nu_0^2+\wt\nu_1^2}\rme^{2\rmi\pi\wt u}+32\rmi\pi q_1^{1/2}\wt\nu_2^2\rme^{-4\rmi\pi\wt u}+16\rmi\pi q_1^{1/4}\wt\nu_2\wt\nu_3\rme^{-2\rmi\pi\wt u},}
which coincides with (\ref{eq:p3_2}) up to a change of arbitrary constants.

\section{Linear problem for the Painlev\'e II equation}
\label{sec:PII}
We construct a linear problem for the Painlev\'e II equation by means of the degeneration process from the linear problem constructed in subsection \ref{sec:4.1}. This process involves substitutions
\eqlb{eq:subII}{\tau_1=\dfrac{t w^2}{2\pi\rmi},\qq\wt u=U\dfrac{w}{2\pi\rmi},\qq z=Z\dfrac w{2\pi\rmi},\qq v=\dfrac Vw,}
scalings of coupling constants, and the limit \eqs{w\rightarrow0}. After applying (\ref{eq:subII}) the canonical Poisson bracket transforms into
\eq{\bfi{V,U}=2\pi\rmi.}

From the decomposition of Lax matrices (\ref{eq:L3}), (\ref{eq:M3}) as series in \eqs{q} one can determine the scalings of coupling constants
\eq{\wt\nu_0=\rmi\dfrac{1+w^2}{4w^3},\qq\wt\nu_1=-\rmi\dfrac{2-w^3}{2w^3},\qq\wt\nu_2= -\rmi\dfrac{1-w^2}{4w^3},\qq\wt\nu_3=\dfrac{\rmi}2\brc{\alpha-1+\dfrac2{w^3}}.}
Since Hamiltonian (\ref{eq:ham3}) for the Painlev\'e III equation diverges as \eqs{w\rightarrow0}, we use the equivalent form (\ref{eq:ham3z}) to obtain the Hamiltonian of the limiting system
\eq{H^{\textrm{II}}=\lim_{w\rightarrow0}\dfrac{w^2}{2\pi\rmi}H^{\mathrm{III}}.}
The Lax matrices of the limiting system are defined as
\eq{L^{\mathrm{II}}=\lim_{w\rightarrow0}w L^{\mathrm{III}},\qq M^{\mathrm{II}}=\lim_{w\rightarrow0}\dfrac{w^2}{2\pi\rmi}M^{\mathrm{III}}.}
Thus, we get the limiting Hamiltonian and the equation of zero curvature in the following form:
\eq{H^{\mathrm{II}}=-\dfrac{\rmi V^2}{2\pi}+\dfrac{\rmi\alpha\brc{U-Z}}{4\pi}+\dfrac{\rmi t\brc{U^2-Z^2}}{8\pi}+\dfrac{\rmi\brc{U^4-Z^4}}{8\pi},}
\eqlb{eq:mo2}{\partial_tL^{\mathrm{II}}-\partial_ZM^{\mathrm{II}}=\bsq{L^{\mathrm{II}},M^{\mathrm{II}}}=\bfi{H^{\mathrm{II}},L^{\mathrm{II}}},}
where
\begin{subequations}
\label{eq:lax2}
\eqlb{eq:p_2L0}{L^{\mathrm{II}}=\brc{\matr{cc}{
V&\dfrac t4-\dfrac{\alpha}Z+\dfrac{Z^2}{16}+\dfrac{UZ}4+\dfrac{U^2}2\\
-\dfrac t4-\dfrac{\alpha}Z-\dfrac{Z^2}{16}+\dfrac{UZ}4-\dfrac{U^2}2&-V
}},}
\eqlb{eq:p_2M0}{M^{\mathrm{II}}=\brc{\matr{cc}{
0&U+\dfrac Z4\\
U-\dfrac Z4&0
}}.}
\end{subequations}
One can rewrite (\ref{eq:mo2}) as the following system of the first-order differential equations:
\eq{\left\{\matr{l}{
\dfrac{\rmd U}{\rmd t}=2V,\svs
\dfrac{\rmd V}{\rmd t}=U^3+\dfrac{t U}2+\dfrac{\alpha}2,
}\right.}
which coincides with the Hamilton equations of motion. Eliminating \eqs{v} from this system we get the Painlev\'e II equation
\eq{\dfrac{\rmd^2 U}{\rmd t^2}=2U^3+tU+\alpha.}

\section{Linear problem for the Painlev\'e I equation}
\label{sec:PI}
In this Section we construct a linear problem for the Painlev\'e I equation via the degeneration process from the linear problem constructed in subsection \ref{sec:4.1}. This process consists of the substitutions
\eqlb{eq:sub1}{\tau_1=\dfrac{w^2 t}{2\pi\rmi},\qq\wt u=U\dfrac w{2\pi\rmi},\qq v=\dfrac Vw,\qq z=-\dfrac12+Z\dfrac w{2\pi\rmi},}
scalings of coupling constants, and the limit \eqs{w\rightarrow0}. After applying (\ref{eq:sub1}) the canonical Poisson bracket transforms into
\eq{\bfi{V,U}=2\pi\rmi.}
The simplest way to determine scalings of coupling constants is to analyze the decomposition of the Hamiltonian (\ref{eq:ham3}) as a series in \eqs{q}. This gives
\eq{\wt\nu_0=\dfrac{\rmi}{2\sqrt2w^{5/2}},\qq\wt\nu_1=-\dfrac{\rmi}{\sqrt2w^{5/2}},  \qq\wt\nu_2-\dfrac1{2\sqrt2w^{5/2}},\qq\wt\nu_3=\dfrac1{\sqrt2w^{5/2}}.}
To obtain the convergent Lax matrices we have to make the following gauge transformation
\eq{L^{\mathrm{III}}\rightarrow gL^{\mathrm{III}}g^{-1},\qq M^{\mathrm{III}}\rightarrow gM^{\mathrm{III}}g^{-1},}
\eq{g=\brc{\matr{cc}{
1&0\\
0&\sqrt w
}},}
and consider the limit
\eq{L^{\mathrm{I}}=\lim_{w\rightarrow0}w gL^{\mathrm{III}}g^{-1},\qq M^{\mathrm{I}}=\lim_{w\rightarrow0}\dfrac{w^2}{2\pi\rmi}gM^{\mathrm{III}}g^{-1}.}
Using the Hamiltonian (\ref{eq:ham3}) for the Painlev\'e III equation we derive the Hamiltonian of the limiting system
\eq{H^{\mathrm{I}}=\lim_{w\rightarrow0}\dfrac{w^2}{2\pi\rmi}H^{\mathrm{III}}=-\dfrac{\rmi V^2}{2\pi}+\dfrac{\rmi tU}{4\pi}+\dfrac{\rmi U^3}{2\pi}.}
After taking the limit the equation of zero curvature becomes
\eqlb{eq:mo1}{\partial_tL^{\mathrm{I}}-\partial_ZM^{\mathrm{I}}=\bsq{L^{\mathrm{I}},M^{\mathrm{I}}}=\bfi{H^{\mathrm{I}},L^{\mathrm{I}}},}
where
\eqlb{eq:lax1}{L^{\mathrm{I}}=\brc{\matr{cc}{
V&\dfrac1{4\sqrt2}\brc{2t+Z^2+2UZ+4U^2}\\
\dfrac1{\sqrt2}\brc{Z-2U}
}},\qq
M^{\mathrm{I}}=\brc{\matr{cc}{
0&\dfrac1{2\sqrt2}\brc{4U+Z}\\
\sqrt2&0
}}.}
One can rewrite (\ref{eq:mo1}) as a system of the first-order differential equations, which is equivalent to the Painlev\'e I equation
\eq{\left\{\matr{l}{
\dfrac{\rmd U}{\rmd t}=2V,\svs
\dfrac{\rmd V}{\rmd t}=3U^2+\dfrac t2
}\right.\qq\Longleftrightarrow\qq
\dfrac{d^2 U}{dt^2}=6U^2+t.}

\section{Conclusion}

We have constructed linear problems for the Painlev\'e equations I--V via the degeneration processes from the linear problem for the Painlev\'e VI equation. These degeneration processes can be described by the following diagram
\eqlb{eq:ds_fin}{\xymatrix{
&\mathrm{PV}\ar[r]&\mathrm{PIV}\\
\mathrm{PVI}\ar[r]\ar[ru]&\mathrm{PIII}\ar[r]\ar[rd]&\mathrm{PII}\\
&&\mathrm{PI}
}}
Thus, we have supplemented the known relations (\ref{eq:pden}) between the Painlev\'e equations with the degeneration scheme (\ref{eq:ds_fin}) for the \eqs{2\times2}-matrix linear problems discussed in this paper.

Since one can obtain the Calogero-Inozemtsev system via the reduction from the \eqs{2\times2} elliptic Schlesinger system with four marked points, it is possible to apply the proposed degeneration process to the general case of the elliptic Schlesinger system. This process can probably give new non-autonomous systems which describe the interaction between the non-autonomous Toda and Calogero-Moser systems. We will study such a degeneration process in the subsequent work.

\section*{Acknowledgements}

We would like to thank M.~A.~Olshanetsky and A.~V.~Zotov for suggesting the problem and many useful discussions.
Both authors have been supported by grants NSh-6941.2012.2,
RFBR 09-02-00-393,
Russian President fund MK-1646.2011.1,
by Rosatom under contract H.4e.45.90.11.1059, and
by Federal Agency of Science and Innovations of Russian Federation under contract 14.740.11.0347.

\appendix

\section{Painlev\'e equations}

In this Section we present a connection between the different forms of the Painlev\'e equations V, IV, and III considered in this paper.
\subsection{Painlev\'e V}
The Painlev\'e V equation has the following rational form:
\eqlb{eq:PVA}{\dfrac{\rmd^2\lambda}{\rmd t^2}=\brc{\dfrac1{2\lambda}+\dfrac1{\lambda-1}}\brc{\dfrac{\rmd\lambda}{\rmd t}}^2-\dfrac1t\dfrac{\rmd\lambda}{\rmd t}+\dfrac{\brc{\lambda-1}^2}{t^2}\brc{\alpha\lambda+\dfrac{\beta}{\lambda}}+\gamma\dfrac{\lambda}t+\delta\dfrac{\lambda\brc{\lambda+1}}{\brc{\lambda-1}}.}
In order to obtain the equivalent form of (\ref{eq:PVA}) which we use in Section \ref{sec:PV} one can perform the following change of variables:
\eq{\lambda\brc{t}=\lambda\brc{u\brc{t}}=-\tg^2\brc{\pi u\brc{t}}.}
As a result, \eqs{\brc{\rmd\lambda/\rmd t}^2} becomes zero.
After the substitution \eqs{t\brc{\tau}=\rme^{\tau}} the derivative \eqs{\rmd\lambda/\rmd t} becomes zero as well, which leads to
\eqlb{eq:PVB}{\dfrac{\rmd^2 u}{\rmd t^2}=\dfrac{\alpha}{2\pi}\dfrac{\sin\brc{\pi u}}{\cos^3\brc{\pi u}}+\dfrac{\beta}{2\pi}\dfrac{\cos\brc{\pi u}}{\sin^3\brc{\pi u}}+2\gamma\rme^{\tau}\sin\brc{2\pi u}+\delta\rme^{2\tau}\sin\brc{4\pi u}.}

\subsection{Painlev\'e IV}
The Painlev\'e IV equation has the following rational form:
\eqlb{eq:PIVA}{\dfrac{\rmd^2\lambda}{\rmd t^2}=\dfrac1{2\lambda}\brc{\dfrac{\rmd\lambda}{\rmd t}}^2+\dfrac32\lambda^3+4t\lambda^2+2\brc{t^2-\alpha}\lambda+\dfrac{\beta}{\lambda}.}
In order to obtain the equivalent form of (\ref{eq:PIVA}) considered in Section \ref{V->IV} one can make the change of variables
\eq{\lambda\brc{t}=u^2\brc{t},}
which leads to
\eqlb{eq:PIVB}{\dfrac{\rmd^2 u}{\rmd t^2}=\dfrac{3u^5}{4}+2tu^3+\brc{t^2-\alpha}u+\dfrac{\beta}{2u^3}.}

\subsection{Painlev\'e III}
The Painlev\'e III equation has the following rational form:
\eqlb{eq:PIIIA}{\dfrac{\rmd^2\lambda}{\rmd t^2}=\dfrac1{\lambda}\brc{\dfrac{\rmd\lambda}{\rmd t}}^2-\dfrac1t\dfrac{\rmd\lambda}{\rmd t}+\dfrac1t\brc{\alpha\lambda^2+\beta}+\gamma\lambda^3+\dfrac{\delta}{\lambda}.}
In order to obtain the equivalent form of (\ref{eq:PIIIA}) which we use in Section \ref{sec:PIII} one can make the change of variables
\eq{\lambda\brc{t}=\rme^{u\brc{t}}.}
Substituting \eqs{t=\rme^{\tau}} into (\ref{eq:PIIIA}) we get
\eqlb{eq:PIIIB}{\dfrac{\rmd^2 u}{\rmd t^2}=\alpha\rme^{\tau+u}+\beta\rme^{\tau-u}+\gamma\rme^{2\brc{\tau+u}}+\delta\rme^{2\brc{\tau-u}}.}

\section{Elliptic functions}
\label{app:ellipticfunctions}

The definitions and properties of elliptic functions used in the paper can be found in \cite{Veyl1} and \cite{Mumford}. The main object is the theta function defined by
\eq{\theta\bsq{\matr{c}{a\\b}}\brc{z,\tau}=\sum_{j\in \mathbb Z}q^{\frac12 (j+a)^2}\mbe\brc{(j+a)(z+b)},}
where \eqs{q=\mbe\brc{\tau}\equiv\exp\brc{2\pi\rmi\tau}}.

We also use the Eisenstein functions
\eq{\varepsilon_k(z)=\lim_{M\rightarrow+\infty}\sum_{n=-M}^M(z+n)^{-k}, \qq k\in \mathbb N,}
\eqlb{eq:eisdef}{E_k(z)=\lim_{M\rightarrow+\infty}\sum_{n=-M}^M\varepsilon_k(z+n\tau).}

To determine limits of Lax matrices we use the following functions:
\eqlb{eq:thetadef}{\vartheta(z)=\theta\bsq{\at{1/2}{1/2}}\brc{z,\tau}=\sum_{j\in\mathbb Z}q^{\frac12\brc{j+\frac12}^2}\mbe\brc{\brc{j+\dfrac12}\brc{z+\dfrac12}},}
\eqlb{eq:phidef}{\phi(u,z)=\dfrac{\vartheta(u+z)\vartheta'(0)}{\vartheta(u)\vartheta(z)},}
\eq{\varphi_{\alpha}\brc{u+\omega_{\beta},z}=\mbe\brc{z\partial_{\tau}\omega_{\alpha}}\phi\brc{u+\omega_{\beta},z},}
\eq{f_{\alpha}\brc{u+\omega_{\beta},z}=
\mbe\brc{z\partial_{\tau}\omega_{\alpha}}\partial_w\phi\brc{w,z}|_{w=u+\omega_{\beta}},}
where \eqs{\omega_{\alpha}=\bfi{0,\frac12,\frac{\tau}2,\frac{1+\tau}2}}.
The functions satisfy the following well-known identities:
\eq{\phi(u,z)\phi(-u,z)=E_2(z)-E_2(u),}
\eqlb{eq:pdphi}{\partial_u\phi(u,z)=\phi(u,z)(E_1(u+z)-E_1(u)),}
parity
\eq{E_k(-z)=(-1)^k E_k(z),}
\eq{\vartheta(-z)=-\vartheta(z),}
\eq{\phi(u,z)=\phi(z,u)=-\phi(-u,-z),}
and quasi-periodicity
\eqlb{eq:quasiperiodicity}{\matr{ll}{
E_1(z+1)=E_1(z),&E_1(z+\tau)=E_1(z)-2\pi \rmi,\\
\\
E_2(z+1)=E_2(z),&E_2(z+\tau)=E_2(z),\\
\\
\vartheta(z+1)=-\vartheta(z),&\vartheta(z+\tau)=-q^{-\frac12}\mbe(-z)\vartheta(z),\\
\\
\phi(u+1,z)=\phi(u,z),&\phi(u+\tau,z)=\mbe(-z)\phi(u,z).}}

\bibliographystyle{unsrt}
\bibliography{references}

\end{document}